# Effects of Inlet and Secondary Flow Conditions on the Flow Field of Rotating Detonation Engines with Film Cooling

Jingtian Yu[1,2], Songbai Yao[1,2,*], Jingzhe Li[1,3], Yihui Huang[1], Chunhai Guo[1,2], Wenwu Zhang[1,2,*]

[1]Zhejiang Key Laboratory of Aero Engine Extreme Manufacturing Technology, Ningbo Institute of Materials Technology and Engineering, Chinese Academy of Sciences, Ningbo 315201, China

[2]University of Chinese Academy of Sciences, Beijing 100049, China

[3]Faculty of Mechanical Engineering and Mechanics, Ningbo University, Ningbo 315211, China

*Corresponding authors: yaosongbai@nimte.ac.cn (S. Yao), zhangwenwu@nimte.ac.cn (W. Zhang)

**Abstract**

A three-dimensional simulation of the rotating detonation engine (RDE) with film cooling is conducted. The aim of this study is to analyze the fluid dynamics and heat transfer of the detonation flow field under the influence of cooling flow from the film holes. Results suggest that when the rotating detonation wave sweeps the film holes, the shape of the wave structure will deform, and the detonation products will invade and block the outflow from the film holes; however, this only occurs temporarily. The structure of the detonation wave will quickly restore to its stable form and, meanwhile, the cooling flow also recovers rapidly and provides adequate protected area on the wall surface and effective thermal protection time in a full propagation cycle of the detonation wave. A parametric analysis indicates that the effective outflow time improves with the increase of the mass flow rate of the cooling flow; on the other hand, the cooling efficiency is more significant downstream from the inlet of the combustor to the outlet. In addition, the thrust and specific impulse of the RDE are also examined under the influence of film cooling.

**Keywords**: Hydrogen detonation; Rotating detonation engine; Film cooling; Numerical simulation

**Nomenclature**

| | |
|---|---|
| RDC | Rotating detonation chamber |
| RDW | Rotating detonation wave |
| RDE | Rotating detonation engine |
| $A$ | Pre-exponential factor |
| $A_o$ | Area of RDC outlet |
| $D$ | Mass diffusion coefficient |
| $d_f$ | Diameter of film hole |
| $d_i$ | Inner diameter of RDC |
| $d_o$ | Outer diameter of RDC |
| $e$ | Internal energy |
| $E_a$ | Activation energy of reaction |
| $F$ | Thrust |
| $g$ | Gravitational constant |
| $h$ | Heat transfer coefficient |
| $H$ | Height of RDW front |
| $I_{sp}$ | Fuel-based specific impulse |
| $I_{tot}$ | Total specific impulse |
| $K$ | Thermal conductivity coefficient |
| $L$ | Length of RDC |
| $\dot{m}_f$ | Mass flow rate of $H_2$ in primary flow |
| $\dot{m}_p$ | Mass flow rate of primary flow |
| $\dot{m}_s$ | Mass flow rate of secondary flow |
| $P$ | Pressure |
| $P_{out}$ | Outlet pressure |
| $P_\infty$ | Ambient pressure |
| $q$ | Chemical energy release |
| $R$ | Gas constant |
| $T$ | Temperature |
| $\overline{T_r}$ | Wall temperature without film cooling |
| $\overline{T_w}$ | Wall temperature with film cooling |
| $T_{w2}$ | Temperature of cooling air |
| $t_p$ | One detonation cycle |
| $t_s$ | Normal cooling film outflow time in one detonation cycle |
| $\eta_t$ | $t_s/t_p$ |
| $\mathbf{U}$ | Velocity |
| $V_e$ | Outlet velocity |
| $v_d$ | Average detonation velocity |
| $Y$ | Species mass fraction |
| $\theta$ | Efficiency of film cooling |
| $\boldsymbol{\tau}$ | Viscous stress tensor |
| $\dot{\omega}$ | Chemical reaction rate |
| $\rho$ | Density |

## 1. Introduction

Characterized by pressure gain combustion (PGC), detonation-based engines are acclaimed to be thermodynamically more efficient than conventional engines based on a constant pressure cycle [1, 2]. As an innovative scheme, propulsion systems are expected to be benefited considerably from the advance of detonation-based propulsion technologies in terms of fuel consumption, manufacturing and pollution control [1], and in particular, a proliferation of studies have been made in the field of rotating detonation engine (RDEs).

Recent developments in the RDE have heightened its potential integration with gas turbines [3-6], ramjets [7, 8], rocket engines [9, 10], etc. For example, Walters et al. [11] investigated the characteristics and maintainability of rotating detonations with methane-air fuel at wide operating conditions where the mass flow rates range from 200-500 $kg/m^2 \cdot s$ and the equivalence ratios were 0.85-1.2. The research was aimed for land-based power generation systems and it was found that the characteristic exhaust velocity of the RDE could reach 97.5-99.5% of the theoretical value compared to the thermodynamic cycle of constant-volume combustion. In addition, Journell et al. [12] analyzed the combustion process of the RDE operating at working conditions similar to gas turbines by means of chemiluminescence and particle image velocimetry (PIV). Goto et al. [13] conducted a rocket sled test for the thrust performance assessment of an RDE with an inner diameter of 60.5 mm. Thrust was generated for 2 seconds and a specific impulse of 144 s was measured. The same research group have recently conducted a flight test of a sounding rocket in which the RDE was incorporated in what they called a detonation engine system and worked for 6 seconds during the flight. Zhou et al. [14] analyzed the ignition initiation process of the rotating detonation wave in the combustion chamber integrated with turbines and several types of propagation modes were observed at different equivalence ratios. Wu et al. [15] studied the interactions between the rotating detonation wave and turbine blade and examined the pressure oscillation phenomena. A spinning pulsed detonation mode was reported by Shen et al. [16] when they investigated the operation of RDEs at elevated back-pressure induced by the integration with gas turbines. On the other hand, ramjet RDEs at different flight conditions have also been widely examined. For example, Meng et al. [17] conducted experiments at an approximate Mach 4 and 20 km altitude condition, and Frolov et al. [18] tested the ramjet RDE at an air steam of Mach 5.7 and stagnation temperature of 1500 K. RDE fueled by liquid propellants or clean energy are also attracting increasing attention and led to numerous studies in which kerosene [19-22], JP-10 [23], n-heptane [24-26], ammonia [27], coal-hydrogen-air mixture [28] are all considered. New design of RDC configurations was also proposed, i.e., in the studies of Betelin et al. [29] and Mikhalchenko and Nikitin [30], fuel mixture was injected from both the inlet and side walls.

However, as pointed out in the review article of Roy et al. [1], the durability of detonation-based propulsion systems was regarded as one of the most challenging issues for engineering applications. The detonation wave is a high-frequency shock coupled with rapid chemical reactions, which can cause thermal deformations on the crucial components, i.e., the rotating detonation chamber (RDC). In the absence of any means of thermal protection, the RDC is subject to wear and tear within seconds of operation. Bykovski et al. [31] measured the heat fluxes at the inner and outer wall surfaces of the RDE and the maximum heat flux was observed in the mixing region, and it was reported by Zhou et al. [32] that the highest heat flux occurred was located at a distance equal to 20% of the length of the RDC. Randall et al. [33] investigated the heat transfer between the inner and outer walls in their numerical studies and the results showed that the inner wall was subjected to much higher thermal loads than the outer wall. Also, heat flux at the downstream of the combustion chamber was smaller than the heat flux near the front of the rotating detonation wave (RDW). By means of $H_2O$ absorption spectroscopy, Rein et al. [34] measured the average and transient temperatures inside the channel of the RDE, confirming that the temperatures varied considerably along the axial direction.

Some cooling schemes have been considered for the thermal protection of RDEs. For example, Ishihara et al. [35] analyzed the variation of the heat flux on the chamber walls made with C/C composites. The RDE operated for 10.2 s continuously at a total mass flow rate of 96 g/s and an equivalence ratio of 1.63. Claflin et al. [36] achieved 20 s of continuously rotating detonations using the water-cooling approach. Theuerkauf et al. [37] measured the transient temperature on the combustion chamber walls using a platinum resistance temperature sensor. They designed convection cooling channels for both the inner and outer walls of the RDE where water was used as the coolant, and successfully stable rotating detonations.

As an effective cooling technique, film cooling has been widely used for thermal protection of turbines, combustors, nozzles, and other crucial components of propulsive systems [38]. However, there has been very limited research on the application of film cooling on the RDE. Ota et al. [39] and Goto et al. [40] placed a series of crosswise-arranged on the bottom plate and outer wall of the RDE where oxygen and ethylene would enter and cooling was achieved under the influence of convection. Cooling film would form on the outer wall of the RDE and they managed to control the outer wall temperature within 850 K; however, the possible effects on the structure and propagation of the RDW were not reported in details. In a numerical study of Tian et al. [41], the interactions between the secondary flow from film cooling and the RDW was investigated. It was found that the secondary flow would oscillate periodically and the cooling effect was most significant on the downstream region swept by the oblique shock wave. Given the very distinct characteristics of detonation combustion compared with deflagration, it is of necessity to obtain an in-depth understanding of the fluid dynamics, heat transfer, and cooling efficiency of the secondary flow in the rotating detonation field under various working conditions; also, the effect of film cooling on the propulsive performance of the RDE should be taken into consideration. Therefore, in this study numerical simulations are conducted to analyze comprehensively the effective protected area, interactions between the RDW and cooling flow, and thrust performance under various inlet and secondary flow conditions. Meanwhile, in consideration of the rapid and continuous rotation of the RDW in the combustor, a characteristics parameter is proposed to represent the effective protection time that film cooling can offer in a complete cycle of the RDW as complementary to the conventionally used cooling effectiveness parameter.

## 2. Numerical methods

### 2.1 Physical modeling

The three-dimensional reactive Navier-Stokes equations are solved for the simulations, i.e.,

$$\frac{\partial \rho}{\partial t} + \nabla \cdot (\rho \mathbf{U}) = 0 \quad (1)$$

$$\frac{\partial (\rho \mathbf{U})}{\partial t} + \nabla \cdot (\rho \mathbf{U}\mathbf{U}) + \nabla p = \nabla \cdot \boldsymbol{\tau} \quad (2)$$

$$\frac{\partial (\rho e)}{\partial t} + \nabla \cdot \big((\rho e + P)\mathbf{U}\big) = \nabla(\boldsymbol{\tau} \cdot \mathbf{U}) + \nabla \cdot (K\nabla T) - \rho q \dot{\omega} \quad (3)$$

$$\frac{\partial (\rho Y)}{\partial t} + \nabla \cdot (\rho Y \mathbf{U}) + \nabla(\rho D \nabla Y) - \rho \dot{\omega} = 0 \quad (4)$$

Here $P$ is the pressure, $T$ is the temperature, $\rho$ is the density, $\mathbf{U}$ is the velocity vector, $\boldsymbol{\tau}$ is the viscous stress tensor, and $e$ is the internal energy. The thermal conduction and mass diffusion coefficients are termed as $K$ and $D$. The chemical reaction of hydrogen/air is modeled by a one-step mechanism formulated by the Arrhenius kinetics,

$$\frac{dY}{dt} = \dot{\omega} = -A\rho Y \exp\left(-\frac{E_a}{RT}\right) \quad (5)$$

where $A = 9.87 \times 10^8 \text{ s}^{-1}$ is the pre-exponential factor, $E_a = 3.1 \times 10^7$ J/kmol is the activation energy of reaction, and $R$ is the gas constant.

The turbulence is described by the realizable k–ε turbulence model. Compared with the standard and RNG k–ε turbulence models, the realizable k–ε turbulence model is found to be more proper for separated flows as well as flows with complex secondary flow features. The non-equilibrium wall functions are selected to handle the near-wall flow, which is particularly suitable when there are large pressure gradients. The simulations are carried out using ANSYS Fluent in which the density-based solver is used and the second-order upwind scheme is adopted.

The physical model for the simulations is a rotating detonation chamber (RDC) with a channel depth of 4 mm (see Fig. 1(a)). The diameters of the inner wall ($d_i$) and outer wall ($d_o$) are 32 mm and 40 mm, respectively, and the length of chamber ($L$) is 40 mm. In practice, multiple cooling holes could be arranged along the axial direction of the RDC for full thermal protection. However, in consideration of the symmetrical structure of the RDC and, more importantly, the computational efficiency, the arrangement is simplified to a single row of cooling holes, i.e., eight oblique cylindrical film cooling holes with a diameter ($d_f$) of 0.8 mm are spaced equally along the length of the combustor with an interval of 4 mm. The first film hole is placed 3.2 mm away from the inlet. The dimensions of the film holes are annotated in Fig. 1(c). The film hole leans at a 30-degree angle ($\alpha$) to the axial direction of the RDC according to the experimental studies on film cooling, such as refs. [42, 43]. The number of cooling holes is decided according to the axial dimension of the RDC. We ensure that the cooling holes are spaced at sufficient distances so that they can provide efficient thermal protection along the length of the RDC. The computation domains, including the RDC and the film holes, are discretized by structured mesh with hexahedral elements (Fig. 1(b)) to guarantee sufficient resolution of the flow field around the film holes.

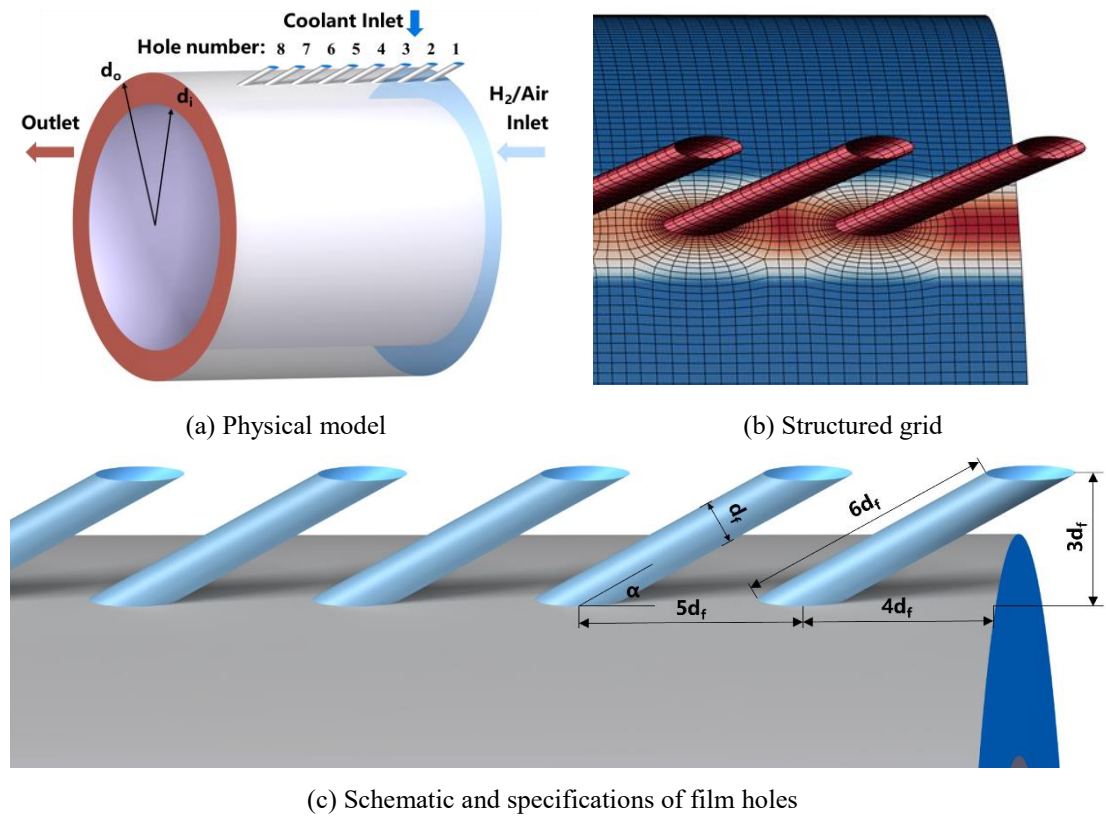

(a) Physical model

(b) Structured grid

(c) Schematic and specifications of film holes

**Figure 1.** Physical model and mesh grid of the RDC.

Mass flow inlet conditions are enforced on both the inlet flow of the RDC (termed as the primary flow) and the cooling secondary flow of the film holes. The primary flow is a stochiometric hydrogen-air mixture and the mass flow rates ($\dot{m}_p$) range from 100 g/s to 200 g/s. On the other hand, the cooling flow is ambient air at the mass flow rates ($\dot{m}_s$) of 0.4 - 1.6 g/s. Pressure outlet and no-slip flow walls are implemented for the boundary conditions and, meanwhile, on the surface of the chamber walls the heat transfer coefficient is estimated to $h = 15 \text{ W}/(\text{m}^2 \cdot \text{K})$ according to the typical range of $h$ for heat transfer by free air convection [44], and the surface emissivity $\varepsilon = 0.9$. The RDE is initially filled premixed hydrogen-air mixture and ignited by a hot spot at $P = 2$ MPa and $T = 2000$ K, a very common approach for initialization in RDE simulations, and a single-wave mode RDE will be obtained. It is possible for the RDE to work in a multi-wave mode [45-52], but this type of scenario is not considered in the current study. The mass flow rate conditions for the primary flow and film cooling are summarized in Table 1.

| Case | Mass flow rate of primary flow ($\dot{m}_f$, g/s) | Mass flow rate of cooling flow ($\dot{m}_s$, g/s) | Primary inlet | | | |
|---|---|---|---|---|---|---|
| | | | T | P | $Y_{H_2}$ | $Y_{O_2}$ |
| 1 | 100 | 0.8 | 300 K | 0.1 MPa | 0.02875 | 0.23 |
| 2 | 150 | 0.8 | | | | |
| 3 | 200 | 0.8 | Secondary inlet | | | |
| 4 | 100 | 0.4 | T | P | $Y_{O_2}$ | |
| 5 | 100 | 1.2 | 300 K | 0.1 MPa | 0.23 | |
| 6 | 100 | 1.6 | | | | |
| 7 | 100 | 0* | Outlet | | | |
| 8 | 150 | 0* | T | | P | |
| 9 | 200 | 0* | 300 K | | 0.05 MPa | |

* Without film cooling

**Table 1.** Boundary and working conditions.

**2.2 Grid sensitivity analysis**

We have conducted the simulations at three different resolutions with a total number of 1.03 million, 0.67 million, and 0.46 million cells to check the grid sensitivity of the numerical methods and setup. The results are shown in Fig. 2 where the flow fields are at the quasi-steady state. It can be seen that the main structure of the RDW and the flow field around the film hole are found to be almost the same. We also place pressure monitors on the three flow fields to record the pressure traces of the RDWs with time and results are shown in Fig. 3. The computed detonation velocities are found to be close, which are 1981.6 m/s, 1964.5 m/s, and 1921.7 m/s for the meshes with 0.46, 0.67, and 1.03 million cells, respectively, all of which are very close to that given by NASA's CEA program [53] under the same initial conditions. The mesh size used by the simulation case in Fig. 2(b) will be adopted in the remainder of the present study in which the averaged mesh sizes in the axial, radial and azimuthal directions are 0.36 mm, 0.33 mm, and 0.29 mm, respectively.

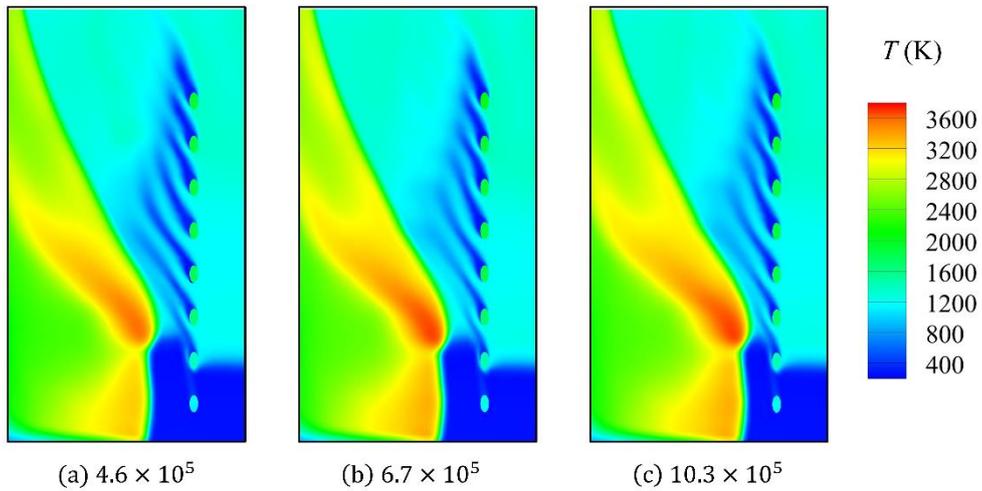

(a) $4.6 \times 10^5$    (b) $6.7 \times 10^5$    (c) $10.3 \times 10^5$

**Figure 2.** Temperature contours of the flow fields at the three different resolutions.

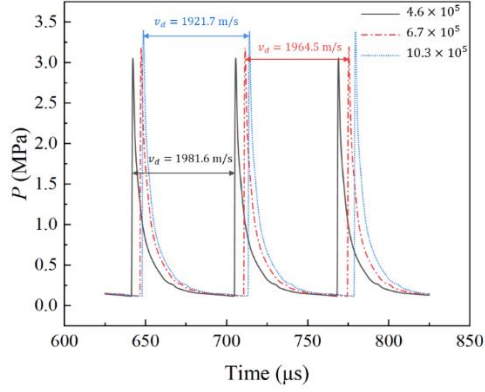

**Figure 3.** Pressure-time profiles of the RDWs at the three different resolutions.

Smirnov et al. [54] proposed a method to evaluate the accumulation error for simulations of reacting flows. In our simulations, the RDW begins to reach a stable state after approximately 400 μs, and the estimated accumulation error is about $S_{err} = 3.6\%$ using this approach; The RDW will continue to propagate for more cycles until t = 1000 μs, and the accumulate error becomes $S_{err} = 5.6\%$, still very close to the recommended range of 1%-5% in ref. [54].

## 3. Results and discussion
### 3.1 Main features of the RDW flow with film cooling

The film cooling holes are activated simultaneously when the RDE is ignited. As shown in Fig. 4, it can be seen that after half cycle of rotation, the RDW will encounter the film holes. The high-temperature detonation products will come into the film holes and for the time being there is no cooling air flowing out, thus the secondary flow at low temperatures has no effect on the detonation wave. The high-temperature products in the film holes will interact with the secondary flow and raise its temperature for some certain time. For the film holes downstream, similar phenomena occur under the influence of the oblique shock wave (OSW). The interactions between the RDW and the secondary flow will also slightly change the structure of the RDW, but the recovery is very fast and after about 1/8 rotating cycle the RDW restores to its original stable shape.

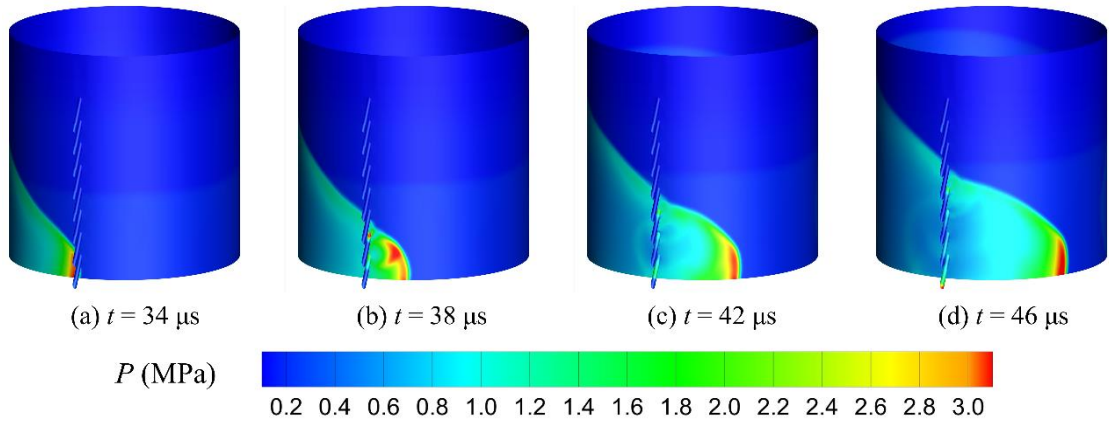

**Figure 4.** Pressure distributions after ignition. Film cooling holes are activated after ignition.

Fig. 5 illustrates the flow field of the RDW passing through the film hole during stable operation. At this time a self-sustained RDW has been obtained and stably run for 9 cycles in a quasi-steady state. It will be used as the benchmark for analysis. Before the detonation wave reaches the position of the film hole, it can be seen that the secondary flows of the eight film holes come out normally. So far, the secondary flow has no noticeable effect on the structure and propagation of the RDW. The interaction mainly occurs after detonation waves pass through film holes. When the RDW passes through the film hole, the flow direction of the low-temperature secondary flow changes instantaneously towards the propagation direction of the RDW. In turn, the secondary flow will also cause disturbance to the RDW and OSW, as can be seen from the temperature distributions. However, since the velocity of detonation wave is several times faster than that of the secondary flow, the interruptions caused by the secondary flow are not significant, and the RDW continues to propagate stably. Therefore, in what follows, we will focus on the thermal protection of film cooling on the RDW flow.

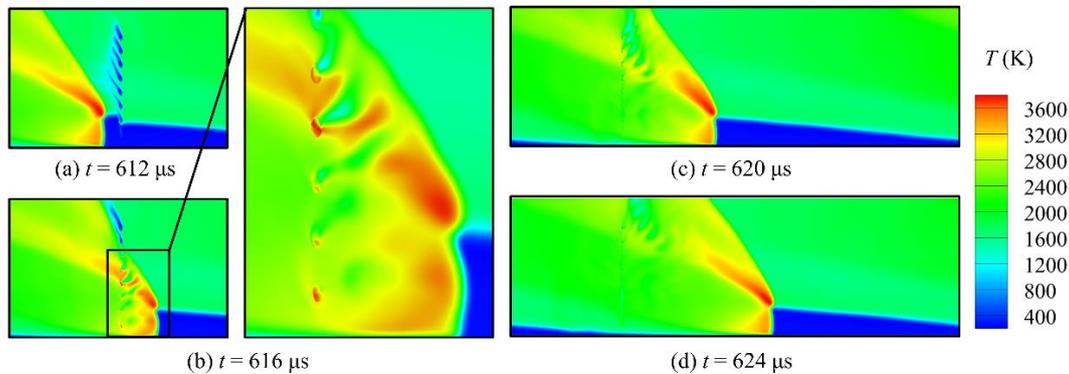

**Figure 5.** Temperature distributions and stable propagation of the RDW in the flow field with film cooling (612μs after the detonation ignition).

### 3.2 Thermal protection performance of film cooling

In Fig. 6, seven snapshots of the $O_2$ mass fraction iso-surfaces (colored by temperature) are plotted in a time sequence indicated by the arrow symbol, which means that the RDW propagates from left to right. In the present study, the mass flow rates of the cooling flow range from $\dot{m}_s = $ 0.4 g/s -1.6 g/s and the outflow conditions during a complete cycle of the RDW are shown. The RDW will interact with the low temperature secondary flow before it reaches the film hole. The temperature of the secondary flow swept by the detonation wave rises instantaneously and the flow direction changes under the influence of the detonation high-speed flow field. The secondary flow heated by the RDW will break off from the film hole and form a cluster of gas mixture with irregular shape. The gas mixture outflowed from each film hole is either connected together or separated, and eventually all of them will be exhausted toward the outlet with the burned products.

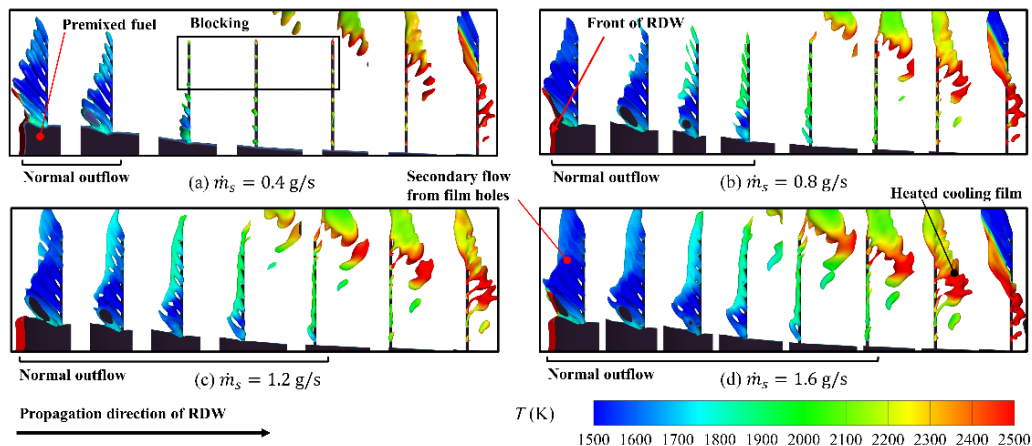

**Figure 6.** Outflow from the film holes at different mass flow rates. (a) $\dot{m}_s = $ 0.4 g/s, (b) $\dot{m}_s = $ 0.8 g/s, (c) $\dot{m}_s = $ 1.2 g/s, (d) $\dot{m}_s = $ 1.6 g/s.

It can be seen that when the RDW flows through the film hole, some high-temperature and high-pressure detonation products will invade the film hole and heat up the low temperature secondary flow inside. However, as the cooling air gradually outflows from the film hole under the driven pressure, a balance of pressure can be achieved and the cooling flow is able to outflow normally and continuously. At $\dot{m}_s = 0.4$ g/s, the film holes will be blocked by the RDW for a longer period of time; with the rise of the secondary flow mass flow rate, it becomes easier to reach a pressure balance. According to Fig. 6 (b) and Fig. 6 (c), the period of time when the film holes are blocked by the RDW is significantly reduced after $\dot{m}_s$ rises to 0.8 g/s -1.2 g/s. In the case of $\dot{m}_s$= 1.6 g/s, after the detonation wave sweeps through the film hole, the cooling flow rapidly recovers and the cooling flow is able to outflow normally in almost the complete rotational cycle of the RDW.

A comparison between the cases in Fig. 6 also shows the variation of the area covered by the cooling flow with the increase of $\dot{m}_s$. The extension of the protected area is evidently reflected in the axial direction. In the azimuthal direction, the velocity of the cooling flow is not comparable to that of the RDW, and thus the increase of $\dot{m}_p$ does not bring considerable difference of the protected area in this direction. Secondly, a higher mass flow rate of the cooling flow also promotes the interactions between the outflow from each film hole, and thus will improve the protection of the overall domain of the RDC.

Additionally, the increase of $\dot{m}_s$ will alleviate the blocking phenomenon caused by the invasion of the detonation produces into the film hole. In order to quantitatively analyze the normal outflow time, the ratio of the normal outflow time to the full cycle of the RDW, $\eta_t$, is introduced here as a characteristic parameter to represent the proportion of time that the secondary flow can normally flow out during the stable rotating detonation combustion time. Let $t_p$ represent the time of one detonation cycle and $t_s$ represent the time of normal outflow from the film hole in one detonation cycle, $\eta_t$ is defined as

$$\eta_t = \frac{t_s}{t_p}. \tag{6}$$

For better accuracy, the calculations of $t_s$ and $t_p$ are averaged based on multiple cycles.

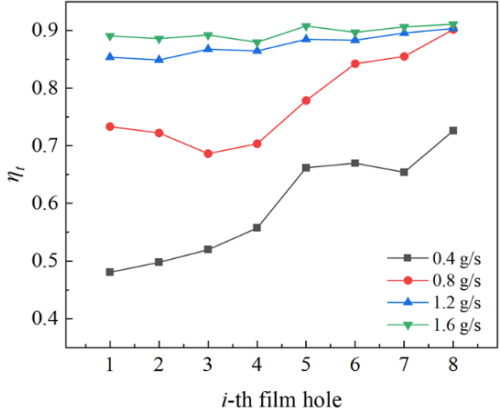
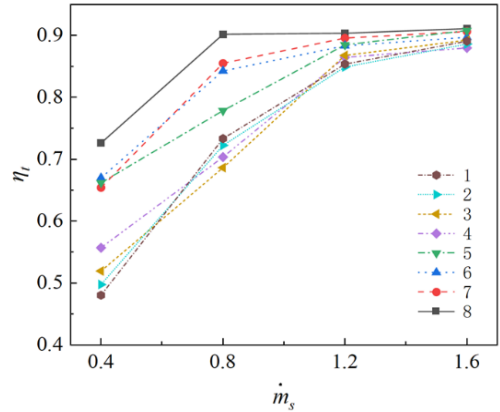

(a) Variation of $\eta_t$ along axial location  (b) Variation of $\eta_t$ with $\dot{m}_s$

[a]**Figure 7.** $\eta_t$ of the film holes and variations with the mass flow rates of the cooling flow.

Eight monitoring points are arranged at the outlet center of the film holes to measure the outflow (normal) velocity in which the sign of the velocity can indicate whether the film hole is blocked or not, and the results are shown in Fig. 7(a). The trends are in good agreement with the observations in Fig. 6, that is, when $\dot{m}_s$ is 0.4 g/s, $\eta_t$ ranges between 50% and 70%, which means that during one detonation cycle more than 40% of the time the film holes are blocked. In this case, the effective protection time is much shorter and cannot offer adequate thermal protection. As $\dot{m}_s$ increases, $\eta_t$ demonstrates an obvious upward trend. When the mass flow rate is increased to 0.8 g/s, $\eta_t$ is significantly improved; it can be increased to more than 85% at 1.2 g/s. A further increase of $\dot{m}_s$ to 1.6 g/s does not introduce considerable improvement of $\eta_t$, i.e., it can be observed from Fig. 7(b) that the increasing trend of $\eta_t$ slows down and stabilizes at about 90%. The possible reason is that the velocity of the RDW is one order of magnitude higher than that of the secondary flow, thus the high-speed and high-pressure detonation wave and detonation products will inevitably invade the film hole after passing through it. Increasing the mass flow rate of the secondary flow can alleviate the required time to reach the pressure balance and shorten the blocking time, but it cannot completely eliminate such phenomenon.

---

[a] In the published version of Figure 7, scatter points were mistakenly plotted in reverse order for labels 1-8. Here is the corrected version.

In analogy to the measurements of film cooling effectiveness for gas turbine engines [55, 56], the following parameter is defined for a quantitative analysis of the efficiency of film cooling,

$$\theta = \frac{\overline{T_r} - \overline{T_w}}{\overline{T_r} - T_{w2}} \quad (7)$$

where $\overline{T_w}$ and $\overline{T_r}$ refer to the wall temperatures with and without film cooling. $T_{w2}$ is the temperature of cooling air flowing into the inlet of film hole. Among the cases, the value of $T_{w2}$ is 300 K. $\overline{T_w}$ and $\overline{T_r}$ are given by the temperatures averaged on multiple rotation cycles of the RDW.

A series of monitoring points are arranged 1 mm downstream of the edge of each film hole in which the radial position of the monitoring point is 0.5 mm inward from the outer wall. These monitoring points are numbered in the same manner as the film holes. Figure 8 shows the film cooling efficiency $\theta$ of the eight monitoring points at $\dot{m}_s$ of 0.8 g/s -1.6 g/s.

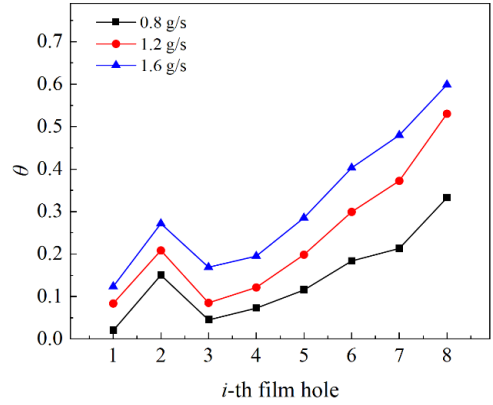

**Figure 8.** The film cooling efficiency $\eta_r$ of the film holes for cases 1, 5 and 6.

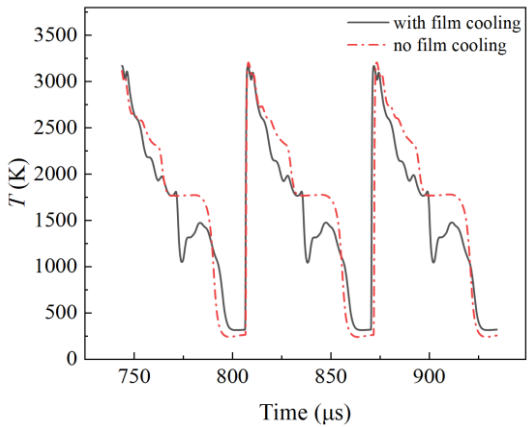

(a) 1st film hole

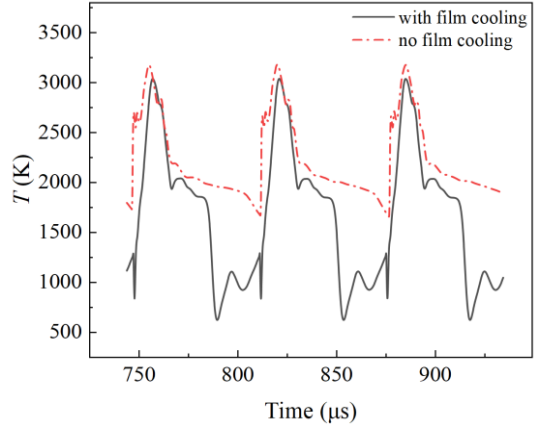

(b) 6th film hole

**Figure 9.** Temperature profiles with and without film cooling (cases 1 and 7) at the monitoring locations of the 1st and 6th film holes near the RDW and OSW regions.

In general, it shows that the cooling efficiency tends to increase downstream along the RDC, but there are differences of heat transfer between the film holes. Here we present a detailed analysis of the cooling holes as they are placed at different locations where the flow field conditions are considerably different. For the first cooling hole, it is located within the height of the RDW, i.e., the fuel refill zone where the temperature is only about 300 K. The fresh fuel at low temperatures can already provide some cooling effect., thus it is actually not very necessary to place a cooling hole within the height of the RDW, the cooling efficiency is not significant.

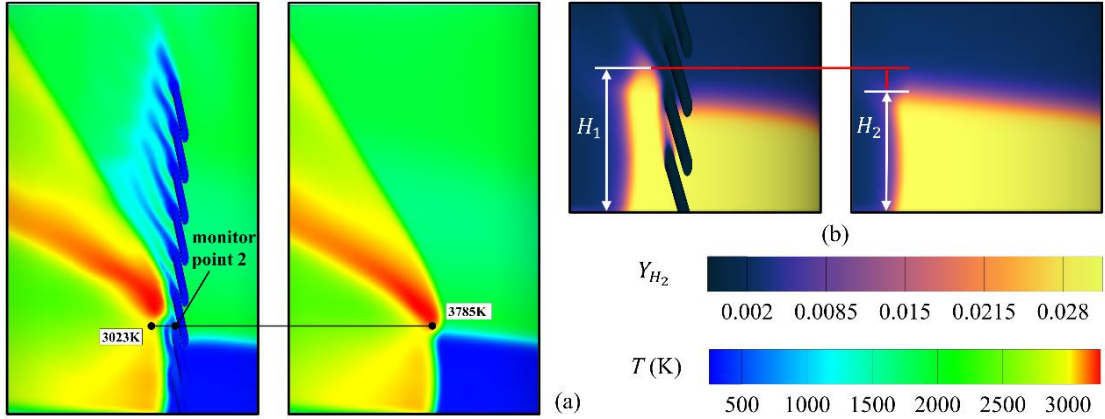

**Figure 10.** Temperature and $H_2$ mass fraction contours near second film hole: (a) Temperature (b) Mass fraction of $H_2$.

On the other hand, the second film hole is located at the junction of the RDW and the OSW. At different mass flow rates of the cooling flow, the cooling efficiency of the second film hole is better than that of its two neighboring ones. An explanation is given here. As shown in Fig. 10, without film cooling, this position is located at the downstream of the junction of the RDW and OSW where the local temperature is high. The flow direction of the low-temperature fuel mixture from the inlet is consistent with the axial component of the outflow direction of the cooling flow. The low-temperature fuel upstream of the second film hole will be brought downstream by the cooling flow under the influence of the outflow from the first film hole. The height of the fresh fuel, which is also the height of the RDW, will therefore be slightly higher ($H_1$ and $H_2$ in Fig. 10(b)). Under the combined influence of the local temperature distribution and the cooling effect of fresh fuel, the cooling efficiency is higher at the second film hole.

The other film holes (nos. 3-8) are located in the region where the OSW passes. The average temperature increases in the axial direction, but the highest transient temperature decreases, and $\theta$ increases in the downstream direction. Figure 9 shows the temperature-time curve of the monitoring points at the $1^s$ and $6^{th}$ film holes at the affected zones of the RDW and OSW. With film cooling, the peak temperature of the monitoring point decreases, and the temperature drops faster after the RDW sweeps.

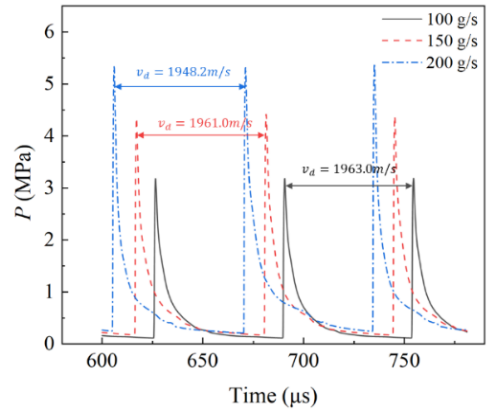

**Figure 11.** Time-pressure curves of cases 1, 2, 3 with film cooling.

An analysis is also carried out to investigate the performance of the cooling flow

(mass flow rate is fixed) under different inlet conditions of the RDE. First, we examine the peak pressure of the RDW at different fuel mass flow rates, which will affect the outflow from the film hole since a pressure balance is required for the cooling flow to come out normally. It can be observed from Fig. 11 that the peak detonation pressure increases significantly with the increase of the primary mass flow rate. Meanwhile, Fig. 12 shows the mass fraction of oxygen as an indicator of the cooling flow in one complete cycle of the RDW. Under different mass flow rates of the fuel mixture, the variation of the average detonation velocity is very small and the coverage area of film cooling in the tangential direction does not change much, but the outflow rate of the film hole significantly decreases. Therefore, as shown in Fig. 13, with the rise of the primary mass flow rate, the normal outflow time (in ratio) $\eta_t$ decreases. At the mass flow rate of $\dot{m}_p = 200$ g/s, $\eta_t$ ranges between 55% and 65% and the blocking time occupies more than 40% in one rotational detonation cycle. Compared with the case of 100 g/s, $\eta_t$ decreases more than 25% on average. Therefore, in order to achieve enhanced thermal protection effect, the mass flow rate of the secondary flow $\dot{m}_s$ should increase with $\dot{m}_p$.

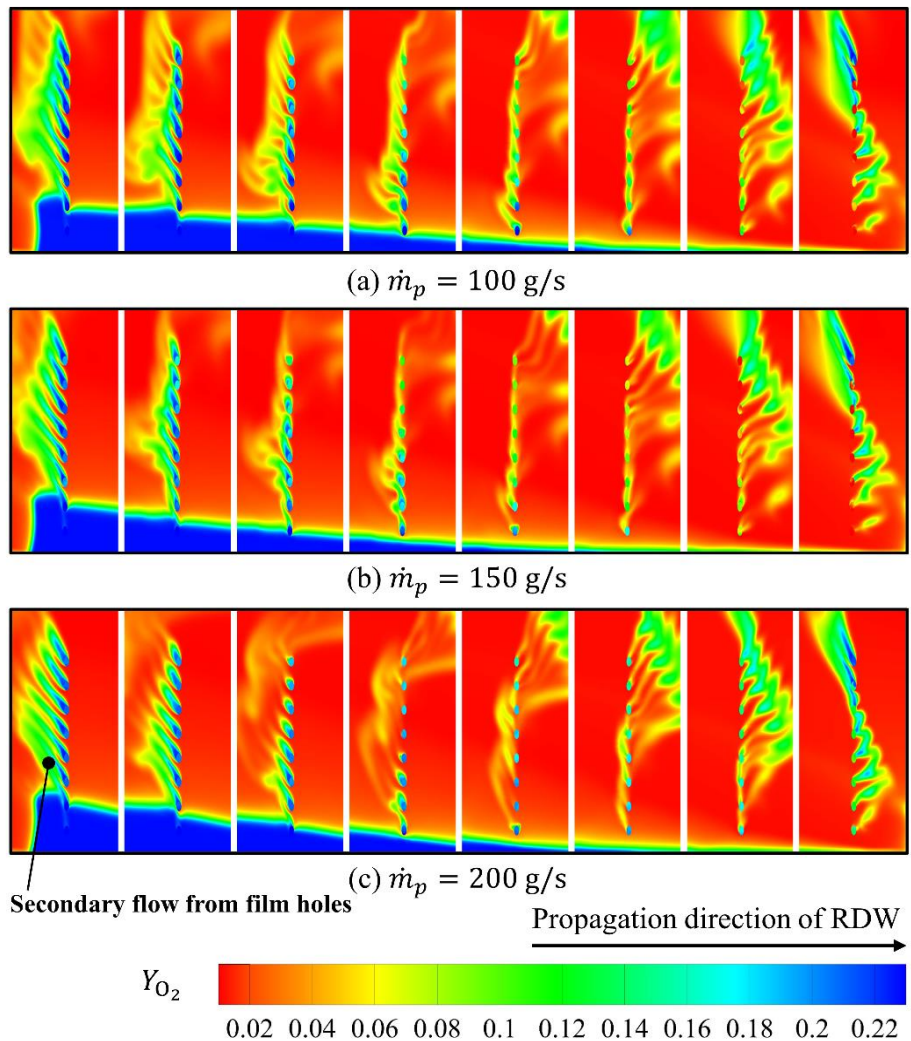

(a) $\dot{m}_p = 100$ g/s

(b) $\dot{m}_p = 150$ g/s

(c) $\dot{m}_p = 200$ g/s

Secondary flow from film holes

Propagation direction of RDW

$Y_{O_2}$  0.02 0.04 0.06 0.08 0.1 0.12 0.14 0.16 0.18 0.2 0.22

**Figure 12.** Outflow conditions of the film holes at various fuel mass flow rates ($\dot{m}_p$).

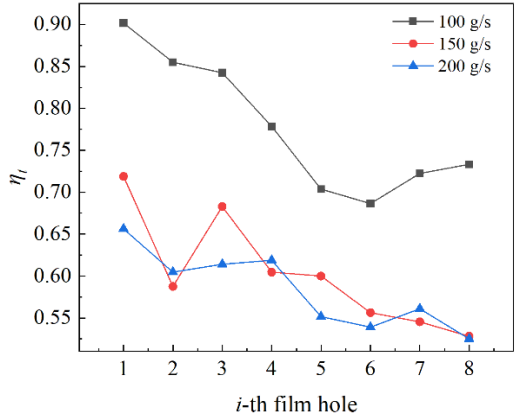

**Figure 13.** Normal outflow time (in ratio) $\eta_t$ of the film holes (cases 1, 2, 3).

## 3.3 Propulsive performance of the RDE with film cooling

Here we will focus on the effects of film cooling on the performance of the RDE. In the analysis of the propulsion performance, the thrust and specific impulse are of great concern, which are defined as follows [57]:

$$F(t) = \int_{outlet} [\rho V_e^2 + P_{out} - P_\infty] dA_o \quad (8)$$

where $V_e$ the outlet velocity, $P_{out}$ is the outlet pressure, $P_\infty$ is the ambient pressure. Given that there are secondary flows from the film holes, the total specific impulse is also computed, that is,

$$I_{sp}^{tot}(t) = \frac{F(t)}{\dot{m}_{tot} g} \quad (9)$$

where $\dot{m}_{tot}$ is the total mass flow rate, $\dot{m}_{tot} = \dot{m}_p + \dot{m}_s$.

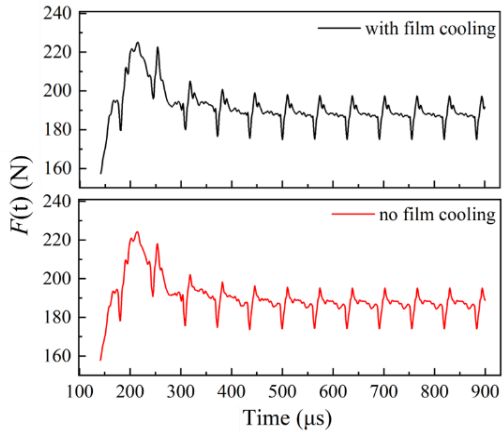

**Figure 14.** Time-thrust curves of the cases with and without film cooling (case 1 and case 7).

In Fig. 14, the thrust-time curves of case 1 (with film cooling) and case 7 (without film cooling) are shown. It can be seen that after about 450 microseconds from the ignition, the RDW has reaches the stable stage, and the data afterwards are collected for analysis.

In Fig. 15, the variation of the average thrust $\bar{F}$ and average total specific impulse $\bar{I}_{sp}^{tot}$ with the increase of the fuel mass flow rate is presented for the cases without and with film cooling; for the latter scenario, the mass flow rate of the cooling flow is fixed at $\dot{m}_s = 0.8$ g/s. It can be seen the difference of the thrust and specific impulse between the cases without and with film cooling is minor, indicating that the reactions of the RDW are not significantly affected by the cooling flow. This is because the propagation velocity of the detonation wave is orders of magnitude faster than that of the outflow from the film holes and, in the present study, the film cooling flow maintains at a low mass flow rate as there is only one single-row of film holes along the axial direction. When the mass flow rate of the cooling flow $\dot{m}_s$ increase from 0.8 g/s to 1.6 g/s (see Fig. 16), while the $\dot{m}_p$ is fixed, it can be seen that the additional flow from film cooling will make contributions to the total mass flow rate on the outlet, thus a slight increase of the thrust, but the total specific impulse tends to decline. We also compute the fuel specific impulse based on Eq . (9). Since $\dot{m}_s$ is not taken into consideration, the average fuel-based specific impulse $\bar{I}_{sp}$ for the cases with film cooling are both at the magnitude of 6600 s, which is close to the experimental measurements for hydrogen-air RDEs [58].

13 / 19

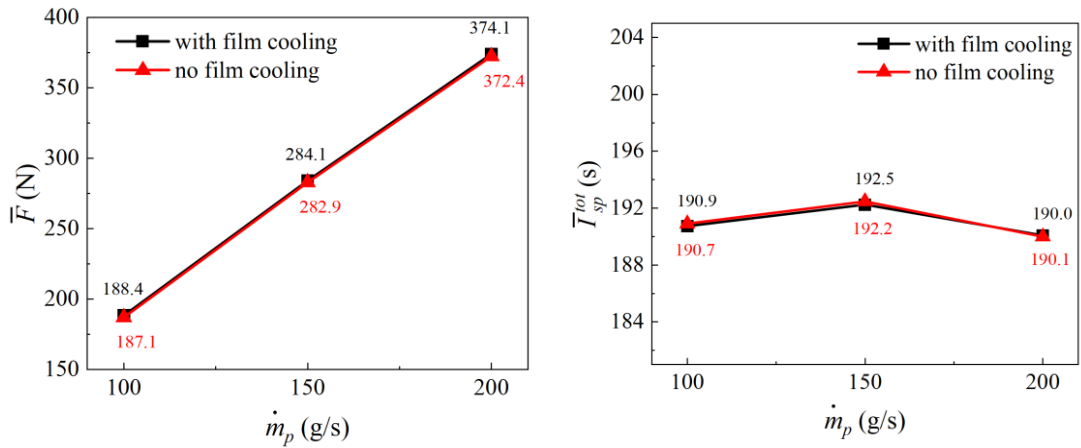

**Figure 15.** Variation of the average thrust and total specific impulse with and without film cooling.

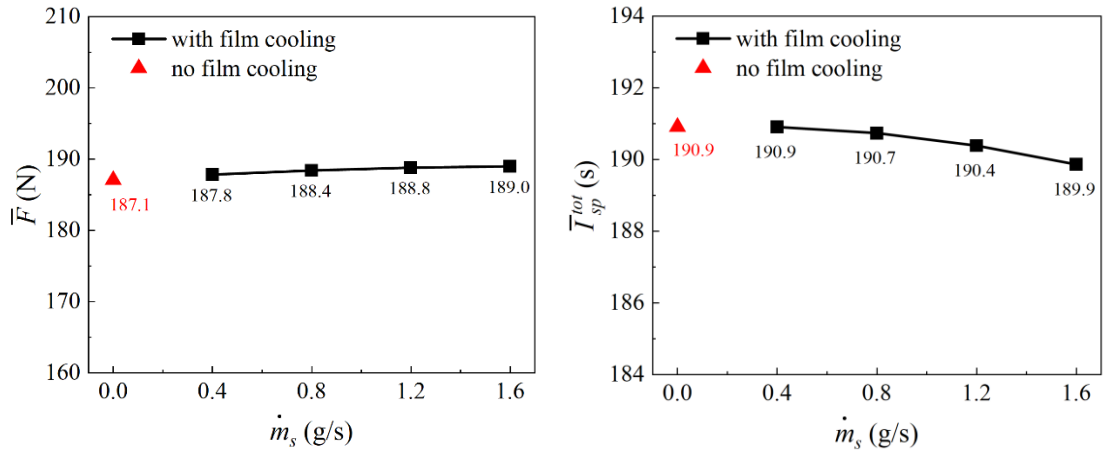

**Figure 16.** Variation of the average thrust and total specific impulse with $\dot{m}_s$.

The temperature distributions of the outlet section of the RDC at different mass flow rates of $\dot{m}_s$ are shown in Fig. 17. Under the influence of cooling flow, the local temperature of the outlet section near the film holes is smaller. And with the increase of the mass flow rate of the secondary flow, the heat transfer between the secondary flow and the primary flow are more significant, thus a larger low-temperature zones can be seen in the outlet section in comparison with the situation without film cooling.

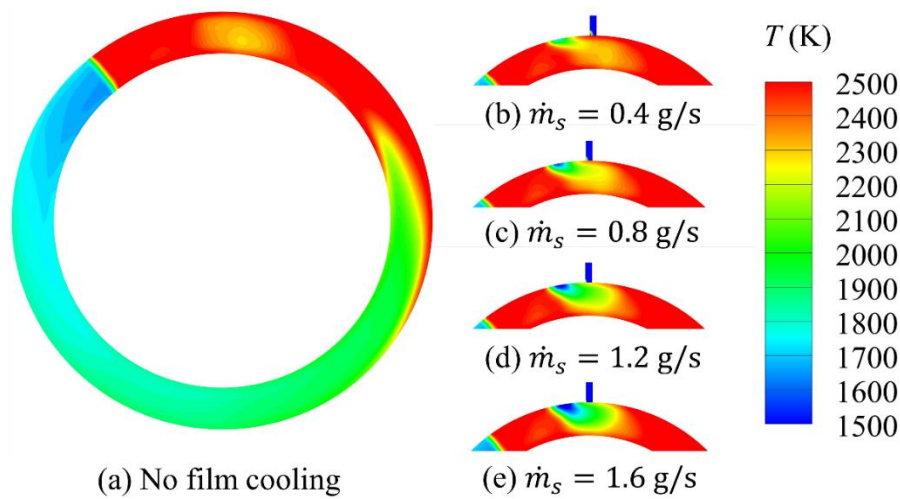

**Figure 17.** RDE outlet temperature distributions as $\dot{m}_s$ ranges from 0.4 – 1.6 g/s.

Finally, we would like to note that in the present study, the RDE is fueled by a premixed stochiometric hydrogen-air mixture. In case of non-premixed RDEs, the process of fuel and oxidizer mixing should be considered and the effect of film cooling on the mixing process should be carefully investigated. Also, in real applications, the cooling holes should be arranged along not just the axial but also the azimuthal direction of the RDC. In this case, the mass flow rate of cooling flow will be larger and the adverse effect on the specific impulse will be more pronounced, an issue that requires further investigation.

## 4. Conclusions

In this numerical study, stable rotating detonations are achieved in the RDC with film cooling. Results show that the secondary cooling flow can deform the RDW for a very short period of time, but the cooling flow does not cause quenching of the detonation and self-sustained RDW can be maintained. The main findings are summarized as follows:

1. When the mass flow rate of the film cooling flow is small, i.e., $\dot{m}_s$ = 0.4 g/s, the blocking time of the film holes due to the invasion of detonation products is long and there is substantial backflow. This blocking phenomenon is significantly reduced after $\dot{m}_s$ increases to 0.8 g/s -1.6 g/s.
2. A higher mass flow rate of the cooling flow will increase of the protected area on the wall surface, but mainly along the axial and radial directions. In the azimuthal direction, the flow velocity is dominated by the propagation velocity of the RDW. The film cooling efficiency gradually increases along the axial direction of the RDC and can be up to 0.6. Also, the normal outflow time, represented by the ratio of the normal outflow time to a complete propagation cycle of the RDW, can reach 90% at high mass flow rates of $\dot{m}_s$.
3. Overall, the film cooling flow does not cause considerable decline of the propulsive performance of the RDE. The average thrust and specific impulse remain almost the same. Due to the contributions of momentum from the secondary flow, the thrust will slightly increase with the mass flow rate of $\dot{m}_s$ when the mass flow rate of the premixed fuel mixture remains the same, but the total specific impulse will decrease.